\documentclass[aps,pra,twocolumn,groupedaddress,showpacs]{revtex4}

\begin{document}

\title{
Quantum entanglement of identical particles}

\author{Yu Shi}

\email[Email address:]{ys219@phy.cam.ac.uk}

\affiliation{Department of Applied Mathematics and Theoretical Physics,
University of Cambridge, Wilberforce Road,
 Cambridge CB3 0WA, United Kingdom}

\affiliation{Theory of Condensed Matter, Cavendish Laboratory,
University of Cambridge,
Cambridge CB3 0HE, United Kingdom}

%(Phys. Rev. A)

\begin{abstract}
We consider  entanglement in a system of fixed number  of
identical particles. Since any operation should be symmetrized
over all the identical particles and there is the precondition
that the spatial wave functions overlap, the meaning of
identical-particle entanglement  is fundamentally different from
that of distinguishable particles. The  identical-particle
counterpart of the Schmidt basis is shown to be the
single-particle basis in which the one-particle reduced density
matrix is diagonal. But it does not play a special role in the
issue of entanglement, which depends on the single-particle basis
chosen. The nonfactorization due to (anti)symmetrization is
naturally excluded by using the (anti)symmetrized basis or,
equivalently, the particle number representation. The natural
degrees of freedom in quantifying the identical-particle
entanglement in a chosen single-particle basis are occupation
numbers of different single particle basis states. The
entanglement between effectively distinguishable spins is shown to
be a special case of the occupation-number entanglement.
\end{abstract}


\maketitle

How does one characterize  entanglement in a fixed number of identical
particles?
Obviously,  a correct characterization must exclude the
nonfactorization due to (anti)symmetrization. Here, we clarify
that it can be done by using the (anti)symmetrized basis, which is
equivalent to the particle number representation. This naturally
leads to the use of occupation numbers of different
single-particle basis states as the (distinguishable) degrees of
freedom in quantifying identical-particle entanglement {\em even
when the number of particles is conserved}. The occupation-numbers
of different modes have already been used in quantum
computing~\cite{milburn}.  The use of modes was made in a previous
study of identical-particle entanglement, based on formally
mapping the Fock space to the state space of qubits or harmonic
oscillators~\cite{zanardi}, but it was under the unphysical
presumption of full access to the Fock space. We shall elaborate
that the concept of entanglement in a system of identical
particles is fundamentally different from that of distinguishable
particles, for which entanglement is invariant under local unitary
transformations. There is no local operation that  acts only on
one of the identical particles.  The single-particle basis
transformation is made on each particle and chooses a different
set of particles in representing the many-particle system. Thus,
the entanglement property of a system of identical particles
depends on the {\em single-particle} basis  used. The particle
number basis state {\em for a fixed number of particles} is just
the normalized (anti)symmetrized basis in the configuration space,
i.e., Slater determinants or permanents. Therefore,  the
occupation-number entanglement in a fixed number of particles is
nothing but the situation that the state is a superposition of
different Slater determinants or permanents. Another consequence
is that the two-identical-particle counterpart of the Schmidt
decomposition, which we call Yang decomposition since the
corresponding transformation of an antisymmetric matrix was first
obtained by Yang long ago~\cite{yang}, does not play a similar
role in characterizing the entanglement. On the other hand, we
show that like the Schmidt basis, the Yang basis is the
single-particle basis in which the one-particle reduced density
matrix is diagonal. It is a common wisdom to treat the
entanglement between spins of identical particles, when they are
effectively distinguished in terms of another degree of freedom,
in the way of distinguishable particles. We show that it is in
fact a special case of the occupation-number entanglement with a
constraint on the accessible subspace of the Fock space.

In terms of the product basis $|k_1,\cdots,k_N\rangle \equiv
|k_1\rangle \otimes\cdots\otimes |k_N\rangle $,   the $N$-particle
state is
\begin{equation}
|\psi\rangle = \sum_{k_1,\cdots, k_N}
 q(k_1,\cdots,k_N)|k_1,\cdots,k_N\rangle, \label{p}
\end{equation}
where the  summations are  made over $k_1,\cdots,k_N$
independently, the coefficients $q(k_1,\cdots,k_N)$ are
(anti)symmetric.

It is often  convenient to use the unnormalized (anti)symmetrized
basis, $|k_1,\cdots,k_N\rangle^{(\pm)}=\sum_P^{N!}
(-1)^{P}|k_1,\cdots,k_N\rangle$, where $P$ denotes permutations,
 `+' is for bosons while `-' is for
fermions. Suppose that in $k_1,\cdots,k_N$, there are $n_{\alpha}$
$k_i$'s which are $\alpha$, then  there are only
$N!/\prod_{\alpha=0}^{\infty}n_{\alpha}!$ different permutations.
Hence,
$|k_1,\cdots,k_N\rangle^{(\pm)}=\sum'_P(-1)^{P}\prod_{\alpha}
n_{\alpha}|k_1,\cdots,k_N\rangle$, where the summation is only
over all different permutations. The $N$-particle state is then
\begin{equation}
|\psi\rangle = \sum_{(k_1,\cdots, k_N)}
 g(k_1,\cdots,k_N)|k_1,\cdots,k_N\rangle^{(\pm)}, \label{sy}
\end{equation}
where $(k_1,\cdots,k_N)$,
disregarding the order of $k_1,\cdots,k_N$,
is {\em a single index}. Up to the sign depending on the
order of  $k_1,\cdots,k_N$ in $q(k_1,\cdots,k_N)$, $q$ is equal to $g$,
 i.e.,
each set of (anti)symmetrized terms in Eq.~(\ref{p}) corresponds
to one term in Eq.~(\ref{sy}). Equation (\ref{sy}) can be rewritten in
terms of the normalized (anti)symmetrized basis
$|k_1,\cdots,k_N\rangle^{(s)}=\sqrt{1/N!\prod_{\alpha}
n_{\alpha}!} |k_1,\cdots,k_N\rangle^{(\pm)}$, as $|\psi\rangle =
\sum_{(k_1,\cdots, k_N)}
h(k_1,\cdots,k_N)|k_1,\cdots,k_N\rangle^{(s)}$.

For a fixed number of particles, the normalized (anti)symmetrized
basis can be rewritten in terms of the occupation numbers of
different  single-particle basis states. This is  the particle
number representation, in which
\begin{equation}
|\psi\rangle = \sum_{n_1,\cdots, n_{\infty}}
 f(n_1,\cdots,n_{\infty})|n_1,\cdots,n_{\infty}\rangle, \label{pn}
\end{equation}
where  $n_j$ is the occupation number of mode $j$,
 $|n_1,\cdots,n_{\infty}\rangle \equiv
(a_{1}^{\dagger})^{n_1}\cdots
(a_{\infty}^{\dagger})^{n_{\infty}}|0\rangle$, the summations are
subject to the constraint $\sum_{\alpha}n_{\alpha}=N$, hence in
the complete summation, of course most of the $f$'s are zero.

In  Refs.~\cite{schliemann1,schliemann2,schliemann3,you,li}, an
arbitrary $N$-particle state, in an arbitrary single-particle
basis, is inappropriately  written as $\sum_{i_1 \cdots i_N}
w_{i_1\cdots i_N} a_{i_1}^{\dagger}\cdots
a_{i_N}^{\dagger}|0\rangle$, where $w_{i_1\cdots i_N}$ is
(anti)symmetric, and each subscript of the creation operators runs
over all the modes. One should note that the creation or
annihilation operators  are associated with
{\em (anti)symmetrized basis}. For example,
$a_{i}^{\dagger}a_{j}^{\dagger}|0\rangle =\pm
a_{j}^{\dagger}a_{i}^{\dagger}|0\rangle =|1_i\rangle |1_j\rangle
=\frac{1}{\sqrt{2}} (|i\rangle |j\rangle \pm |i\rangle |j\rangle
)$,  where $i \neq j$. Therefore  $a_{i_1}^{\dagger}\cdots
a_{i_N}^{\dagger}|0\rangle$ in these papers may be corrected to
$|i_1\cdots i_N\rangle$.  On the other hand, if one uses the
particle number basis states, {\em no} (anti)symmetrization needs
to made~\cite{al}.

Single-particle basis transformation for identical particles is
{\em not} the counterpart  of the local unitary transformation in
a system of distinguishable particles. It acts on each identical
particle in the same way. For distinguishable particles, local
unitary transformations do not change the entanglement. In
contrast, for identical particles, the entanglement depends on
which single-particle basis is chosen.  Consequently, unlike  the
Schmidt decomposition of distinguishable particles, the Yang
decomposition does not play a special role in identical-particle
entanglement~\cite{real}.

As a simple example,  consider a  two-particle state
$a_{{\mathbf{k}}_1}^{\dagger}a_{{\mathbf{k}}_2}^{\dagger}|0\rangle
\equiv\frac{1}{\sqrt{2}}(|{\mathbf{k}}_1\rangle|{\mathbf{k}}_2\rangle
\pm |{\mathbf{k}}_2\rangle|{\mathbf{k}}_1\rangle)$, assuming
${\mathbf{k}}_1 \neq {\mathbf{k}}_2$. In terms of  momentum basis,
this is only a basis state in  particle number representation.
Written in terms of the  product basis, the non-factorization  is
only due to (anti)symmetrization. Therefore, there is no entanglement.
However,  in terms of the position basis,  it becomes
$\sum_{{\mathbf{r}}_1,{\mathbf{r}}_2} e^{i({\mathbf{k}}_1\cdot
{\mathbf{r}}_1+{\mathbf{k}}_2\cdot{\mathbf{r}}_2)}
a_{{\mathbf{r}}_1}^{\dagger}
a_{{\mathbf{r}}_2}^{\dagger}|0\rangle$, which is entangled.

Since a  single-particle basis transformation is made on every
particle, the entanglement property depends on the single particle 
basis chosen.  
This invalidates the use of the von 
Neumann entropy of the one-particle reduced 
density matrix as a measure of
entanglement for two identical particles~\cite{you}. 
Of course, the von Neumann entropy of any density matrix 
characterizes its decomposition in its eigenbasis, which
is the Yang basis in the case of one-particle reduced density matrix 
for a system of two identical particles, 
as explicitly shown below. The
$n$-particle reduced density matrix for a $N$-particle system is
$\langle k_1',\cdots,k_n'|\rho^{(n)}|k_1,\cdots, k_n \rangle = Tr
(a_{k_1'}\cdots a_{k_n'} \rho a_{k_n}^{\dagger}\cdots
a_{k_1}^{\dagger})$, with $Tr\rho^{(n)}=N(N-1)\cdots(N-n+1)$. One
can find
\begin{equation}
\begin{array}{l}
\langle k_1' \cdots k_n'|\rho^{(n)}|k_1 \cdots k_n\rangle =
\frac{1}{(N-n)!} \times  \nonumber
\\
\sum_{k_{n+1}\cdots k_N} {^{(\pm)}\langle} k_1' \cdots k_n'
k_{n+1}\cdots k_N|\rho|k_1 \cdots k_nk_{n+1}\cdots
k_N\rangle^{(\pm)}.  \label{re}
\end{array}
\end{equation}
For a two-boson  product state
$a_{{\mathbf{k}}_1}^{\dagger}a_{{\mathbf{k}}_2}^{\dagger}|0\rangle$
with ${\mathbf{k}}_1 \neq {\mathbf{k}}_2$, the one-particle
reduced density matrix is given by $\langle
{\mathbf{k}}_1|\rho^{(1)}|{\mathbf{k}}_1\rangle= \langle
{\mathbf{k}}_2|\rho^{(1)}|{\mathbf{k}}_2\rangle=1$ and $\langle
{\mathbf{k}}_1|\rho^{(1)}|{\mathbf{k}}_2\rangle= \langle
{\mathbf{k}}_2|\rho^{(1)}|{\mathbf{k}}_1\rangle=0$, hence the
one-particle partial entropy is $\log 2$, 
contradicting the previous claim.

The dependence of entanglement on the single-particle basis is
consistent with the point of view that individual particles are
excitation of quantum fields, and that each different
single-particle basis, in fact, defines a different set of
particles representing the many-body state. In fact, in many-body
physics, it is  a routine  to make various transformations, which
usually changes the nature of entanglement~\cite{trans}. 

With (anti)symmetrization already made on the basis, the
correlation embedded in the coefficients naturally gives  the
information  on entanglement.
A Slater determinant or permanent is
just a (anti)symmetrized basis state, hence is nonentangled with
respect to the given single-particle basis. 
One can regard a
superposition of Slater determinant or permanent as entangled 
in the given single-particle basis.
Any operation, even a
one-body one, of which the single-particle basis transformation is
an example, acts on all the particles. A transformation from a
superposition of Slater determinant or permanent  to a single
Slater determinant or permanent, in another single-particle basis,
must involve operations on all particles and actually chooses a
different set of particles in representing the state.  In a sense,
there is a builtin nonseparability, based on both the
symmetrization of any operation {\em and the spatial wave function
overlap}. Consistently, without spatial wave function overlap or
under the condition of the so-called remoteness~\cite{peres}, the
symmetrization does not have any physical effect.

Hence, the entanglement is  between different 
single-particle basis states. Whether a certain 
single particle basis state is entangled with other single particle basis 
states can be decided by whether the former is mixed with the latter  
in the single-particle basis transformation which transforms the 
superposition into a single Slater determinant or permanent. 
This can be seen most clearly by using the second quantization. 
For example,  in a two-particle state $1/\sqrt{m}
a_{{\mathbf{k}}_1}^{\dagger}(a_{{\mathbf{k}}_2}^{\dagger}+\cdots+
a_{{\mathbf{k}}_{m+1}}^{\dagger})|0\rangle$, where
${\mathbf{k}}_i$'s are different from each other, $m > 1 $, 
$|{\mathbf{k}}_1\rangle$ state is obviously separated from the others. 
One can obtain the one-particle partial entropy as 
$\log 2 + 1/2 \log m > \log 2$. 

Since,  the distinguishable label is the set of occupation numbers
of different single-particle basis states, clearly they can be
used to quantify entanglement, in the way of entanglement between
distinguishable objects. From Eq.~(\ref{pn}), one obtains the
density matrix as $\langle
n_1',\cdots,n_{\infty}'|\rho|n_1,\cdots,n_{\infty}\rangle =
f^*(n_1',\cdots,n_{\infty}') f(n_1,\cdots,n_{\infty})$, from which
one can obtain the reduced density matrices {\em of occupation
numbers}. For example, the reduced density matrix of mode $1$ is
defined as $\langle n_1'|\rho_1(1)|n_1\rangle
=\sum_{n_2,\cdots,n_{\infty}} \langle
n_1',n_2,\cdots,n_{\infty}|\rho|n_1,n_2\cdots,n_{\infty}\rangle$.
Similarly, the reduced density matrix of the set of modes
$1,\cdots, l$ is
 \begin{equation}
\begin{array}{l}
 \langle n_1',\cdots,n_l'|\rho_l(1\cdots
l)|n_1,\cdots,n_l\rangle =  \nonumber \\
\sum_{n_{l+1},\cdots,n_{\infty}} \langle n_1',\cdots,n_l',
n_{l+1},n_{\infty}|\rho|n_1,\cdots,n_l, n_{l+1},n_{\infty}\rangle,
 \label{lm}
\end{array}
\end{equation}
the nonvanishing elements of which satisfy $\sum_{i=1}^l n_i'=
\sum_{i=1}^l n_i$ as constrained by the particle number
conservation. From these Fock-space reduced density matrices, one
can, for example, calculate  bipartite entanglement between the
occupation numbers of $l$ modes  and the  occupation numbers of
the other modes.

It is important to note that the use of occupation numbers as the
degrees of freedom in characterizing entanglement is valid even
when the particle number is conserved. This physical constraint,
as well as the constraints that for fermions $n_j$ is either $0$
or $1$ and that the number of the relevant modes~\cite{relevant}
may be finite are all automatically satisfied by the set of
nonzero $f's$.  Hence,
this approach is a natural one  within the standard second
quantization formalism, compatible with the representations of the
observables in terms of creation or annihilation operators, which
can be viewed as  coordinated transformations of occupation
numbers of a set of modes. The second-quantized representation of
an $n$-body operator $O$ is $\sum a_{i_1'}^{\dagger}\cdots
a_{i_n'}^{\dagger} \langle i_1'\cdots i_n'|O|i_1\cdots i_n\rangle
a_{i_n}\cdots a_{i_1}$. One can observe that, for example,
 there is no operation which only changes the occupation number
of one mode. One may consider ``second-quantized computation''.

In principle,  one can define  entanglement with respect to any
reference state of the system. In this case, the occupation
number of each mode in defining the relative entanglement  is the
difference with that  in  the reference state, as conveniently
seen by considering the action of
 creation operators.
There are  two reference states  that  are of particular interest.
One is the empty state, as we have
implicitly considered up to now.  Another one
is  the
ground state of the system, which is  suitable when all physical
processes are in  a  same bulk of material. In discussing
the entanglement in a ground state,  it is with
respect to the empty state. The
considerations can  even be  extended to relativistic quantum field
theory, where the ground state is the vacuum.

An important situation  is  that the single-particle basis
includes both spin and  orbit (momentum or position). One can
denote the total index as $({\mathbf{o}},s)$, where ${\mathbf{o}}$
substitutes for momentum $\mathbf{k}$ or position $\mathbf{r}$. A
special case is half-filling, i.e.,  each orbit is constrained to
be occupied by only one particle, i.e. $\sum_{s}
n_{{\mathbf{o}},s}=1$ for each relevant value of $\mathbf{o}$.
Then with the orbit modes  as the labels with which the particles
are effectively distinguished, the entanglement can be viewed as
the spin entanglement among the particles in different orbit
modes. Under the constraint of half filling, the many-particle
state must be a sum of products of
$\prod_{s}|n_{{\mathbf{o}},s}\rangle$, under the constraint
$\sum_{s} n_{{\mathbf{o}},s}=1$, for relevant orbits. In the
many-particle state, one simply rewrites
$\prod_{s}|n_{{\mathbf{o}},s}\rangle$ as $|S_{\mathbf{o}}\rangle$,
where  $S{\mathbf{o}}$ is unambiguously
the $s$ corresponding to $n_{{\mathbf{o}},s}=1$. This rigorously
justifies the common wisdom that although it is meaningless to
identify which particle is in which orbit, it is meaningful to say
that the particle in a certain orbit is spin entangled with the
particle in another orbit. Hence, the entanglement between
Heisenberg spins, which appears as  entanglement between
distinguishable objects, is in fact a special case of
occupation-number entanglement.

Entanglement between Heisenberg spins is the basis of the quantum
computing scheme based on electrons in double quantum
dots~\cite{loss}.  When the electrons are separated in the two
dots, because there are only  one-dot potentials, while the
Coulomb interaction is negligible, the condition of
remoteness~\cite{peres} is satisfied. One can verify that an
antisymmetrization between electrons in different dots has no
physical effects.  On the other hand, when they are close, and the
interaction is appreciable, the antisymmetrization has physical
effects, and  the entanglement can be characterized by using the
full formalism of occupation-number entanglement. During the
interaction period, they access the full Hilbert space, which
includes the state in which the two electrons, with opposite
spins,  locate in a same dot, i.e.
$|1\rangle_{i,\uparrow}\rangle|1_{i,\downarrow}\rangle$, where
$i=1,2$ represents the dots. There are
$4!/2!2!=6$ two-particle antisymmetrized basis states, or
occupation-number basis states. Nevertheless, as far as  the
initial and final states are in single occupancy, and the
Heisenberg model is a valid description, the interaction period
can be viewed as an intermediate process determining the effective
spin coupling, while the leakage into the full Hilbert space of
two identical particles during a two-particle gate operation does
not cause any problem. In terms of the occupation-number states,
the spin state of each electron in each dot is $|\uparrow \rangle
_i= |1\rangle _{i,\uparrow}|0\rangle _{i,\downarrow}$,
$|\downarrow \rangle _i= |0\rangle _{i,\uparrow}|1\rangle
_{i,\downarrow}$. Because a spin qubit is in fact an
occupation-number  state, the loss of identification after
separating from the double occupation, as concerned in
Ref.~\cite{hu}, does not matter.  Note that the intermediate state
with double occupancy is  {\em necessary} for the electrons to
interact in order to undergo a two-qubit operation.

Finally,  we come to the question that what is special about the
Yang basis. For two distinguishable particles, the Schmidt basis
is clearly the one in which  the reduced density matrix of each
particle is diagonal: For $\sum_i c_i|i\rangle_a|i\rangle_b$ of
distinguishable  particles $a$ and $b$,
the elements of the
reduced density matrix of either  $a$ or  $b$ are
given by $\langle i|\rho_{a (b)}|j\rangle = |c_i|^2\delta_{ij}$.
In the following,  we show that the  Yang basis is the basis in which
the one-particle reduced density matrix is diagonal.

In their Yang basis, a two-fermion state is like $|\psi_f\rangle =
c_1(|1\rangle|2\rangle-|2\rangle |1\rangle)+ c_2(|3\rangle
|4\rangle-|3\rangle|4\rangle) +\cdots$ where we use $|i\rangle$ to
denote different single-particle basis
 states.  If $k_2'=k_2$, then $k_1'=k_1$ is necessary for any of $\langle
k_1'k_2'|\rho|k_1k_2\rangle$, $\langle
k_1'k_2'|\rho|k_2k_1\rangle$, $\langle k_2'k_1|\rho|k_1k_2\rangle$
and $\langle k_2'k_1|\rho|k_2k_1\rangle$ to be nonvanishing.
Therefore using Eq.~(\ref{re}), one finds
 $\langle k_1'|\rho^{(1)}|k_1\rangle
=\delta_{k_1'k_1} \sum_{k_2}{^{(-)}\langle}
k_1k_2|\rho|k_1k_2\rangle^{(-)}$. Hence, $\rho^{(1)}$ is diagonal.

In their Yang basis, a two-boson state is like
$|\psi_b\rangle = d_1 |1\rangle |1\rangle + d_2|2\rangle |2\rangle
+\cdots$.
Then one finds
$\langle k_1'k_2'|\rho|k_1k_2\rangle= \langle k_1'k_2'|\rho|k_2'k_1\rangle=
\langle k_2'k_1'|\rho|k_1k_2\rangle = \langle k_2'k_1'|\rho|k_2k_1\rangle
= \delta_{k_1'k_2'} \delta_{k_1k_2}
 \langle k_1k_1|\rho|k_2 k_2\rangle $. Consequently,
 using Eq.~(\ref{re}), one finds
$\langle k_1'|\rho^{(1)}|k_1\rangle =\delta_{k_1'k_1}
{^{(+)}\langle} k_1k_1|\rho|k_1k_1\rangle^{(+)}$. Hence,
$\rho^{(1)}$ is diagonal.

Let us summarize. If one uses the product basis, the coefficients
mix the information on (anti)symmetrization and that on
entanglement. If, instead, the (anti)symmetrization is made on the
basis, then the coefficients unambiguously give the information on
entanglement, with respect to the given single-particle basis.
(Anti)symmetrized basis  is equivalent to particle number
representation, and the  occupation numbers of different modes are
distinguishable degrees of freedom which can be used in
quantifying the entanglement even when particle number is
conserved. Entanglement of identical particles is a property
dependent on which single-particle basis is chosen, as any
operation should act on each identical particle in the same way.
Indeed,  individual particles are excitations of a quantum field,
and the single-particle basis defines which set of particles are
used in representing the many-particle state. 
The many-particle state is entangled in the
corresponding single particle basis when it is not a single Slater
determinant or permanent. The entanglement is between different 
single-particle basis states in the given basis. 
We also show that the entanglement
between effectively distinguishable spins of identical particles
is a special case of the occupation-number entanglement. We have
discussed its use in quantum computing. The (necessary)
leakage into the larger Hilbert space {\em during} the
intermediate two-particle process is harmless. Finally it is shown
that the two-identical-particle counterpart of the Schmidt basis
is the basis in which the one-particle reduced density matrix is
diagonal. In addition to quantum computing implementations
involving identical particles, the result here is also useful for
many-body physics~\cite{shi3}.

This publication is an output from project activity funded by
The Cambridge-MIT Institute Limited.


\begin{thebibliography}{99}
\bibitem{al}
If the creation operators are used, then each coefficient is
defined only up to a set of modes. Nevertheless,  one has the
freedom to split each term. Thus,  in the case of two particles,
the state may be written in terms of a coefficient  matrix and
creation operators. The coefficient matrix may be made to be
(anti)symmetric in consideration of the (anti)commutation
relations of the creation operators. But this is artificial. In
fact, there are infinite number of ways of splitting.
For instance, $a_1^{\dagger}a_2^{\dagger}|0\rangle \equiv
 (xa_1^{\dagger}a_2^{\dagger}\pm
ya_2^{\dagger}a_1^{\dagger})|0\rangle$, where $x+y=1$.
\bibitem{real} The definition of ``quantum
correlation'' and ``Slater rank'' in
Ref.~\cite{schliemann1,schliemann2,schliemann3} is meaningful only
when the given single-particle basis is just the Yang basis. In general,
the proposed ``concurrence'' or the determinant of the coefficient
matrix does not characterize the entanglement property in
the  given single-particle basis. For example, the result on
the two fermions with four-dimensional single-particle Hilbert
space only implies  that iff the determinant is zero, and if one
transforms to the Yang basis, then the state is a single Slater
determinant in the Yang basis.   It does not tell whether it is a
single Slater determinant in the given basis. It is  easy  to find
examples of a state which is a superposition of different Slater
determinants while the determinant of the coefficients vanishes,
e.g.,  $c_{\alpha}|1,2\rangle^{(-)}+c_{\beta}|1,3\rangle^{(-)}$,
where the numbers denote single-particle states, $c_{\alpha}$ and
$c_{\beta}$ are nonzero coefficients.  For two bosons in
four-dimensional single-particle Hilbert space, the vanishing of
the coefficient determinant does not mean single Slater permanent
even in the Yang basis, as claimed in Ref.~\cite{schliemann3}. 
A counterexample is $c_{1}|1\rangle|1\rangle+c_{2}|2\rangle|2\rangle
+c_{3}|3\rangle|3\rangle$, where at least two of the coefficients
are nonzero. Similar problems exist in  those discussions on more
general cases.
\bibitem{trans}  Some
transformations,  e.g.,  the one appearing  in  the exact solution
of independent boson model, are global  canonical transformation
leading  to a diagonalized Hamiltonian. Some transformations, e.g.
that  between Bloch basis and Wannier basis, are single-particle
basis transformations.
 Some transformations, e.g.,  Jordan-Wigner transformation,
 are based on  representing the
original operators in terms of new operators, which makes the
Hamiltonian diagonal and the eigenstates nonentangled. In some
cases, the combination of the original particles gives
quasiparticles, for example, polaron, polariton, exciton, etc.
Some transformations  even  split the original operators, e.g.,
the slave boson approach.
\bibitem{relevant}  Relevant mode  means that in at least
one of  those basis states corresponding to   nonzero $f$, the
occupation number of this mode is nonzero. The occupation-number
state of an irrelevant mode is simply a factor $|0\rangle$ and can
be neglected.
\end{thebibliography}
\end{document}